\documentclass[a4paper,12pt]{article}
\usepackage{epsfig}
\usepackage{amsmath}
\usepackage{amssymb}
\usepackage{amsfonts}
\usepackage{bbm}
\usepackage{fleqn}
\usepackage[footnotesize]{caption}
\usepackage{graphicx}
\usepackage{mathrsfs}
\usepackage[center,footnotesize,hang]{subfigure}

\def\I{\mathrm{i}}

\def\be{\begin{equation}}
\def\ee{\end{equation}}
\def\beq{\begin{equation}}
\def\eeq{\end{equation}}
\def\bea{\begin{eqnarray}}
\def\eea{\end{eqnarray}}

\def\<{\left\langle}
\def\>{\right\rangle}

\allowdisplaybreaks[2]
\begin{document}
\bibliographystyle{OurBibTeX}
\begin{titlepage}
 \vspace*{-15mm}
\begin{flushright}
\end{flushright}
\vspace*{5mm}
\begin{center}
{ \sffamily \LARGE Tri-bimaximal Neutrino Mixing and $\theta_{13}$}
\\[8mm]
S.~F.~King\footnote{E-mail:
\texttt{king@soton.ac.uk}}
\\[3mm]
{\small\it
School of Physics and Astronomy,
University of Southampton,\\
Southampton, SO17 1BJ, U.K.
}\\[1mm]
\end{center}
\vspace*{0.75cm}
\begin{abstract}
\noindent
We propose an extension of tri-bimaximal mixing to include a non-zero reactor
angle $\theta_{13}$ while maintaining the tri-bimaximal predictions for the
atmospheric angle $\theta_{23}=45^o$ and solar angle $\theta_{12}=35^o$.
We show how such tri-bimaximal-reactor mixing can arise at leading order
from the (type I) see-saw mechanism with partially constrained sequential dominance.
Partially constrained sequential dominance
can be realized in GUT models with a non-Abelian discrete
family symmetry, such as $A_4$, spontaneously broken by flavons with a
particular vacuum alignment.
\end{abstract}
\end{titlepage}
\newpage
\setcounter{footnote}{0}
Over the last decade neutrino physics has undergone a revolution
with the measurement of neutrino mass and lepton mixing from
a variety of solar, atmospheric and terrestrial neutrino
oscillation experiments \cite{Mohapatra:2005wg}.
Lepton mixing is described by the $3\times 3$ matrix
\cite{MNS}
\begin{equation}
U =\left(\begin{array}{ccc}
U_{e1} & U_{e2} & U_{e3} \\
U_{\mu1} & U_{\mu2} & U_{\mu3} \\
U_{\tau1} & U_{\tau2} & U_{\tau3} \\
\end{array}\right).
\label{MNS}
\end{equation}
The Particle Data Group (PDG) parameterization of the
lepton mixing matrix (see e.g.\ \cite{PDG}) is:
\begin{eqnarray}\label{Eq:StandardParametrization}
 U = \left(
  \begin{array}{ccc}
  c_{12}c_{13} &
  s_{12}c_{13} & s_{13}e^{-\I\delta}\\
  -c_{23}s_{12}-s_{13}s_{23}c_{12}e^{\I\delta} &
  c_{23}c_{12}-s_{13}s_{23}s_{12}e^{\I\delta}  &
  s_{23}c_{13}\\
  s_{23}s_{12}-s_{13}c_{23}c_{12}e^{\I\delta} &
  -s_{23}c_{12}-s_{13}c_{23}s_{12}e^{\I\delta} &
  c_{23}c_{13}
  \end{array}
  \right) \, P \, ,
\end{eqnarray}
where $s_{13}=\sin \theta_{13}$, $c_{13}=\cos \theta_{13}$
with $\theta_{13}$ being the
reactor angle, $s_{12}=\sin \theta_{12}$, $c_{12}=\cos \theta_{12}$
with $\theta_{12}$ being the
solar angle, $s_{23}=\sin \theta_{23}$, $c_{23}=\cos \theta_{23}$
with $\theta_{23}$ being the
atmospheric angle, $\delta$ is the (Dirac) CP violating phase which is in
principle measurable in neutrino oscillation experiments, and
$P = \mathrm{diag}(e^{\I \tfrac{\alpha_1}{2}},
e^{\I\tfrac{\alpha_2}{2}}, 0)$ contains additional (Majorana)
CP violating phases $\alpha_1, \alpha_2$.

Current global fits for the solar and atmospheric
angles typically allow the following
$1 \sigma$ ranges (in degrees) \cite{Maltoni:2004ei},
\beq
\theta_{12}= 34.5^o\pm 1.4^o, \ \
\theta_{23}= {43.1^o}^{+4.4^o}_{-3.5^o}.
\label{fits}
\eeq
However the situation concerning the reactor angle is more subtle and requires
some brief discussion. In global fits it has been noted that there is a mild
tension between solar data from SNO and KamLAND, and a further weak tension involving
atmospheric data from SuperKamiokande, where both tensions may be
resolved by allowing a non-zero value for $\theta_{13}$, corresponding to
the following value with $1\sigma$ error \cite{Fogli:2008jx},
\beq
\sin^2\theta_{13}=0.016\pm 0.010,
\eeq
corresponding to a $1.2\sigma$ indication of a non-zero value for $\theta_{13}$.
The SuperKamiokande collaboration is presently in the process of analyzing
its SK-I, SK-II and SK-III atmospheric data including the effect of
$\theta_{13}$, so the situation regarding the atmospheric data should be clarified
in the near future. In the meantime, MINOS has announced its first results on
electron appearance \cite{Diwan} and observes an excess of events at the level
of $1.5\sigma$. This may be also interpreted as a hint for $\theta_{13}$,
and this has been combined by Fogli ``at face value''
with the global fit above to give roughly \cite{Fogli:2008jx},
\beq
\sin^2\theta_{13}=0.02\pm 0.01.
\eeq
Thus, apparently following the adage ``many a little makes a mickle'',
one is led to a $2\sigma$ indication for a non-zero value of $\theta_{13}$.
This corresponds to a value for $\theta_{13}$ in the
$1 \sigma$ range (in degrees),
\beq
\theta_{13}= 8^o\pm 2^o.
\label{fits2}
\eeq
In any case it is certainly
theoretically plausible that $\theta_{13}$ could take a value
in the above range \cite{Albright:2006cw}, so it is interesting to consider this possibility,
and we emphasize this more general motivation.

It is well known that the solar and atmospheric data are
consistent with so-called tri-bimaximal (TB) mixing
\cite{HPS},
\begin{eqnarray}
U_{TB} =
\left( \begin{array}{rrr}
\sqrt{\frac{2}{3}}  & \frac{1}{\sqrt{3}} & 0 \\
-\frac{1}{\sqrt{6}}  & \frac{1}{\sqrt{3}} & \frac{1}{\sqrt{2}} \\
\frac{1}{\sqrt{6}}  & -\frac{1}{\sqrt{3}} & \frac{1}{\sqrt{2}}
\end{array}
\right)P,
\label{MNS0}
\end{eqnarray}
corresponding to the mixing angles,
\footnote{Note that different versions of the TB mixing matrix
appear in the literature with the minus signs appearing in
different places corresponding to differing
choices of charged lepton and Majorana phases. We prefer the
convention shown which emerges from the PDG parametrization when the angles
are set equal to those shown in Eq.\ref{TBangles}}
\beq
\theta_{12}= 35.26^o, \ \
\theta_{23}= 45^o, \ \
\theta_{13}= 0^o.
\label{TBangles}
\eeq
The ansatz of TB mixing matrix is interesting due to its symmetry properties which
seem to call for a possibly discrete non-Abelian family symmetry in nature \cite{Harrison:2003aw}.
There has been a considerable amount of theoretical work in this direction
\cite{Chen:2009um,King:2005bj,deMedeirosVarzielas:2005ax,King:2006me,Altarelli:2006kg,Frampton:2004ud}.
The presence of a
non-zero reactor angle as in Eq.\ref{fits2} would be clearly inconsistent with the TB
prediction for the zero reactor angle in Eq.\ref{TBangles}
and so the TB ansatz would be excluded, even though the
predictions for the solar and atmospheric angles remain acceptable.

In this paper we shall explore the possibility of extending the TB mixing
matrix to allow for a non-zero reactor angle $\theta_{13}$,
while at the same time preserving the predictions for the
tri-maximal solar angle and the maximal atmospheric angle
given by Eq.\ref{TBangles}, namely $\theta_{12}= 35.26^o$ and $\theta_{23}= 45^o$.
In order to maintain these predictions requires,
\beq
\frac{|U_{e2}|^2}{|U_{e1}|^2}= \frac{1}{2}, \ \
\frac{|U_{\mu 3}|^2}{|U_{\tau 3}|^2}=1.
\label{TBR0}
\eeq
To leading order in $U_{e3}$ the conditions in Eq.\ref{TBR0} correspond approximately to,
\beq
|U_{e2}|^2 \approx 1/3, \ \ |U_{\mu 3}|^2 \approx 1/2.
\label{TBR}
\eeq
We refer to the above proposal as as tri-bimaximal-reactor (TBR) mixing,
to emphasize that tri-maximal solar mixing and maximal atmospheric
mixing are both preserved while non-zero reactor mixing is introduced.
To leading order in $U_{e3}$ the TBR mixing matrix following from Eq.\ref{TBR} is then,
\begin{eqnarray}
U_{TBR} =
\left( \begin{array}{ccc}
\sqrt{\frac{2}{3}}  & \frac{1}{\sqrt{3}} & U_{e3} \\
-\frac{1}{\sqrt{6}}(1+ \sqrt{2}U_{e3}^*)
& \frac{1}{\sqrt{3}}(1- \frac{1}{\sqrt{2}}U_{e3}^*)
& \frac{1}{\sqrt{2}} \\
\frac{1}{\sqrt{6}}(1- \sqrt{2}U_{e3}^*)
& -\frac{1}{\sqrt{3}}(1+ \frac{1}{\sqrt{2}}U_{e3}^*)
 & \frac{1}{\sqrt{2}}
\end{array}
\right)P.
\label{MNS2}
\end{eqnarray}

The TBR proposal in Eqs.\ref{TBR0}, \ref{TBR} and \ref{MNS2}
should not be confused with the tri-maximal proposal
\cite{Wolfenstein:1978uw} that the second column of the mixing matrix should
consist of a column with all elements equal to $1/\sqrt{3}$, corresponding to the second
neutrino mass eigenstate being a democratic combination of flavour eigenstates,
i.e. $|U_{l2}|^2=1/3$ for $l=e,\mu ,\tau$.
Tri-maximal mixing predicts a solar mixing angle approximately given by $35^o$,
to leading order in $U_{e3}$. However, for $|U_{e3}|=0.14$ (corresponding to $|U_{e3}|^2=0.02$)
there are significant deviations which tend to increase
the prediction for $\theta_{12}$ beyond the $1\sigma$ range quoted in Eq.\ref{fits}
\cite{Grimus:2008tt,Albright:2008rp}. In addition tri-maximal mixing
also predicts deviations from maximal atmospheric mixing
which depend on the combination $U_{e3}\cos \delta$ \cite{Grimus:2008tt,Albright:2008rp}.
For example, according to Fig.2 of \cite{Albright:2008rp},
for $|U_{e3}|=0.15$, tri-maximal mixing would constrain the phase $\delta$ to lie in
one of the two approximate ranges $\delta = \pi /3 - 2\pi /3$ or $\delta = 3\pi /2 - 9\pi /5$
in order to yield an atmospheric angle within its $1\sigma$ experimental range.

In contrast, TBR mixing defined by Eq.\ref{TBR0} predicts both
tri-maximal solar mixing $\theta_{12}= 35^o$ and
maximal atmospheric mixing $\theta_{23}= 45^o$,
for all values of $\theta_{13}$. However, in practice, realistic models only predict
approximate TBR mixing, so the leading order
approximations to TBR mixing in Eqs.\ref{TBR},\ref{MNS2} should provide an adequate approximation.
In addition, as discussed later,
model dependent corrections to the TBR predictions for the solar and atmospheric angles are expected
at some level. Nevertheless, since tri-maximal mixing predicts that
$\tan 2\theta_{23}$ is inversely proportional to $U_{e3}\cos \delta$ \cite{Grimus:2008tt,Albright:2008rp},
while exact TBR mixing predicts maximal atmospheric mixing, these two proposals may be distinguished by
accurate determinations of $\theta_{23}$, $\theta_{13}$ and $\cos \delta$ at future
high precision neutrino facilities.

In general, note that TBR mixing differs from all variants based on preserving a
particular row or column of the TB mixing matrix \cite{Albright:2008rp},
since the approximate TBR requirements $|U_{e2}|^2\approx 1/3$ and $|U_{\mu 3}|^2\approx 1/2$,
together with $U_{e3}$ being non-zero,
results in every row and column of the TBR mixing matrix
being different from that of the TB mixing matrix, as is clear
by comparing Eq.\ref{MNS2} to Eq.\ref{MNS0}.

In the following discussion it is convenient use the expansion about TB mixing
introduced in \cite{King:2007pr},
\be
s_{13} = \frac{r}{\sqrt{2}}, \ \ s_{12} = \frac{1}{\sqrt{3}}(1+s),
\ \ s_{23} = \frac{1}{\sqrt{2}}(1+a),
\label{rsa}
\ee
where the three real parameters
$r,s,a$ describe the deviations of the reactor, solar and
atmospheric angles from their tri-bimaximal values.
In terms of the deviation parameters the $1\sigma$ ranges for the mixing angles
in Eqs.\ref{fits}, \ref{fits2} then translate into
\be
0.14<r<0.24, \ -0.05<s<0.02, \ -0.04<a<0.10,
\label{rsa:ranges}
\ee
with a central value of $r=0.2$. To first order
in $r,s,a$ the lepton mixing matrix can be written as \cite{King:2007pr},
\begin{eqnarray}
U =
\left( \begin{array}{ccc}
\sqrt{\frac{2}{3}}(1-\frac{1}{2}s)  & \frac{1}{\sqrt{3}}(1+s) & \frac{1}{\sqrt{2}}re^{-i\delta } \\
-\frac{1}{\sqrt{6}}(1+s-a + re^{i\delta })  & \frac{1}{\sqrt{3}}(1-\frac{1}{2}s-a- \frac{1}{2}re^{i\delta })
& \frac{1}{\sqrt{2}}(1+a) \\
\frac{1}{\sqrt{6}}(1+s+a- re^{i\delta })  & -\frac{1}{\sqrt{3}}(1-\frac{1}{2}s+a+ \frac{1}{2}re^{i\delta })
 & \frac{1}{\sqrt{2}}(1-a)
\end{array}
\right)P.
\label{MNS1}
\end{eqnarray}
Other related proposals to parametrize the lepton mixing matrix
have been considered in \cite{Li:2004dn}.

In terms of the deviation parameters $r,s,a$, to leading order
the TBR ansatz introduced in Eqs.\ref{TBR},\ref{MNS2}
is equivalent to having $s=a=0$,
corresponding to
\be
s_{13} = \frac{r}{\sqrt{2}}, \ \ s_{12} = \frac{1}{\sqrt{3}},
\ \ s_{23} = \frac{1}{\sqrt{2}},
\label{rsa1}
\ee
and a corresponding leading order mixing matrix,
\begin{eqnarray}
U_{TBR} =
\left( \begin{array}{ccc}
\sqrt{\frac{2}{3}}  & \frac{1}{\sqrt{3}} & \frac{1}{\sqrt{2}}re^{-i\delta } \\
-\frac{1}{\sqrt{6}}(1+ re^{i\delta })  & \frac{1}{\sqrt{3}}(1- \frac{1}{2}re^{i\delta })
& \frac{1}{\sqrt{2}} \\
\frac{1}{\sqrt{6}}(1- re^{i\delta })  & -\frac{1}{\sqrt{3}}(1+ \frac{1}{2}re^{i\delta })
 & \frac{1}{\sqrt{2}}
\end{array}
\right)P,
\label{MNS3}
\end{eqnarray}
which is equivalent to Eq.\ref{MNS2}, with
$|U_{e3}|=0.14$ corresponding to $r=0.2$.
By contrast, to leading order in $r$, the tri-maximal proposal \cite{Wolfenstein:1978uw,Grimus:2008tt,Albright:2008rp}
corresponds to having $s=0$ and $a=-(r/2)\cos \delta$, which for $r=0.2$ would imply,
$a=-0.1\cos \delta$.

Having postulated the TBR form of the lepton mixing matrix, the next question is what
neutrino mass matrix does this correspond to?
In the flavour basis, in which the charged lepton mass matrix is
diagonal and the TBR arises from the neutrino sector, the effective neutrino
mass matrix $({M^{\nu}_{eff}})^{TBR}$ is given in terms of the neutrino masses
$m_1,m_2,m_3$ by,
\bea
({M^{\nu}_{eff}})^{TBR} & = &
U_{\mbox{\scriptsize TBR}}{\rm diag}
(m_{1}, \; m_{2}, \; m_{3})U_{\mbox{\scriptsize TBR}}^{T} \nonumber \\
& = &
m_{1} \Phi_{1}\Phi_{1}^{T} + m_{2} \Phi_{2}\Phi_{2}^{T} + m_{3} \Phi_{3}\Phi_{3}^{T} ,
\label{MTBR}
\eea
where we have written the mixing matrix in terms of three column vectors
\beq
U_{\mbox{\scriptsize TBR}}=(\Phi_1,\Phi_2,\Phi_3).
\label{columns}
\eeq
It is clear that each column of the TBR
mixing matrix in Eq.\ref{MNS3} can be written as a sum of a column of the TB mixing matrix
in Eq.\ref{MNS0} plus a correction proportional to the reactor parameter $r$.
Then, to first order in $r$, we find that the matrices in
Eq.\ref{MTBR} can be written as a sum of the TB matrices plus a correction
proportional to the reactor parameter $r$,
\bea\label{MTBR2}
\Phi_{1} \Phi_{1}^{T} & = &  \frac{1}{6}
\left(\begin{array}{rrr}
4 & -2 & 2 \\
-2 & 1 & -1 \\
2 & -1 & 1
\end{array}\right)
- \frac{1}{3}re^{i\delta }
\left(\begin{array}{rrr}
0 & 1 & 1 \\
1 & -1 & 0 \\
1 & 0 & 1
\end{array}\right), \nonumber \\
\Phi_{2} \Phi_{2}^{T} & = &  \frac{1}{3}
\left(\begin{array}{rrr}
1 & 1 & -1 \\
1 & 1 & -1 \\
-1 & -1 & 1
\end{array}\right)
- \frac{1}{6}re^{i\delta }
\left(\begin{array}{rrr}
0 & 1 & 1 \\
1 & 2 & 0 \\
1 & 0 & -2
\end{array}\right)
, \nonumber \\
\Phi_{3}\Phi_{3}^{T} & = &  \frac{1}{2}\left(\begin{array}{rrr}
0 & 0 & 0 \\
0 & 1 & 1 \\
0 & 1 & 1
\end{array}\right)
+ \frac{1}{2}re^{-i\delta }
\left(\begin{array}{rrr}
0 & 1 & 1 \\
1 & 0 & 0 \\
1 & 0 & 0
\end{array}\right).
\eea

From above we may write $({M^{\nu}_{eff}})^{TBR}$ as the symmetric matrix,
\begin{equation}\label{eq:csd-tbm2}
({M^{\nu}_{eff}})^{TBR}=
\left(\begin{array}{ccc}
a & b & c \\
. & d & e \\
. & . & f
\end{array}\right),
\end{equation}
where,
\begin{eqnarray}\label{abc}
a &=& \frac{2}{3}m_1+\frac{1}{3}m_2,\nonumber \\
b &=& -\frac{1}{3}m_1+\frac{1}{3}m_2
-re^{i\delta }\left(\frac{1}{3}m_1+\frac{1}{6}m_2\right)
+re^{-i\delta }\left(\frac{1}{2}m_3\right)
,\nonumber \\
c &=& \frac{1}{3}m_1-\frac{1}{3}m_2
-re^{i\delta }\left(\frac{1}{3}m_1+\frac{1}{6}m_2\right)
+re^{-i\delta }\left(\frac{1}{2}m_3\right)
,\nonumber \\
d &=& \frac{1}{6}m_1+\frac{1}{3}m_2 +\frac{1}{2}m_3
+re^{i\delta }\left(\frac{1}{3}m_1-\frac{1}{3}m_2\right)
,\nonumber \\
f &=& \frac{1}{6}m_1+\frac{1}{3}m_2 +\frac{1}{2}m_3
+re^{i\delta }\left(\frac{1}{3}m_1+\frac{1}{3}m_2\right),\nonumber \\
e &=& -\frac{1}{6}m_1-\frac{1}{3}m_2 +\frac{1}{2}m_3.
\end{eqnarray}
In the limit that $r=0$, $({M^{\nu}_{eff}})^{TBR}$ reduces to the TB neutrino mass matrix
$({M^{\nu}_{eff}})^{TB}$, and the relations $b=-c$ and $d=f$ and $-e = a+b-d$ emerge
as the characteristic signatures of the TB neutrino mass matrix in the flavour basis,
in the convention for the TB matrix in Eq.\ref{MNS0}. This implies that the origin
of the reactor parameter
$r$ is due to a violation of the family symmetry that would lead to TB mixing.

We now show how TBR mixing can arise at leading order from see-saw models based on
sequential dominance (SD) \cite{King:1998jw}.
To set the notation, recall that,
in the type I see-saw mechanism, the starting point is a heavy right-handed Majorana
neutrino mass matrix
$M_{RR}$ and a Dirac neutrino mass matrix (in the left-right convention) $M_{D}$, with
the light effective left-handed Majorana
neutrino mass matrix $M_{eff}^{\nu}$ given by the type I see-saw formula
\cite{Minkowski:1977sc},
\begin{equation}\label{eq:meff}
M_{eff}^{\nu} = M_{D} M_{RR}^{-1} M_{D}^{T}.
\end{equation}
In a basis in which $M_{RR}$ is diagonal, we may write,
\begin{equation}
M_{RR} = \mbox{diag}(M_A, M_B, M_C)
\end{equation}
and $M_{D}$ may be written in terms of three general column vectors $A,B,C$,
\begin{equation}
M_{D} = (A,B,C).
\end{equation}
The see-saw formula then gives,
\begin{equation}
\label{eq:seesawmeff}
M_{eff}^{\nu} =
 \frac{AA^{T}}{M_A} + \frac{BB^{T}}{M_{B}} +\frac{CC^{T}}{M_{C}}.
\end{equation}
As noted in \cite{Chen:2009um} $({M^{\nu}_{eff}})^{TB}$ may be achieved if
\begin{equation}
\label{FD2}
A = \frac{a}{\sqrt{6}}
\left(
\begin{array}{r}
2 \\
-1 \\
1
\end{array}
\right),\ \
B = \frac{b}{\sqrt{3}}
\left(
\begin{array}{r}
1 \\
1 \\
-1
\end{array}
\right),\ \
C  = \frac{c}{\sqrt{2}}
\left(
\begin{array}{r}
0 \\
1 \\
1
\end{array}
\right),
\end{equation}
which was referred to as form dominance (FD), since
the constraints in Eq.\ref{FD2} lead to a form
diagonalizable $({M^{\nu}_{eff}})^{TB}$
diagonalized by $U_{\mbox{\scriptsize TB}}$ (in this basis) with physical neutrino
mass eigenvalues given by $m_1=a^2/M_A$, $m_2=b^2/M_B$, $m_3=c^2/M_C$.
It is interesting to compare FD to
Constrained Sequential Dominance (CSD) defined in
\cite{King:2005bj}. In CSD a strong hierarchy $|m_1|\ll |m_2| < |m_3|$ is assumed
which enables $m_1$ to be effectively ignored (typically this is achieved by
taking $M_A$ to be very heavy leading to a very light $m_1$) then CSD is defined by
only assuming the second and third conditions in Eq.\ref{FD2} \cite{King:2005bj}.
Thus CSD is seen to be just a special case of FD corresponding to a strong neutrino mass
hierarchy. FD on the other hand is more general
and allows any choice of neutrino masses including
a mild hierarchy, an inverted hierarchy or a quasi-degenerate mass pattern.

Clearly to achieve TBR mixing we must relax one or more of the conditions in
Eq.\ref{FD2}. There are many ways to do this, and here we just consider one simple
example. Let us begin by considering
a strong neutrino mass hierarchy $|m_1|\ll |m_2| < |m_3|$ corresponding
to $M_A$ being very heavy leading to a very light $m_1$ as in CSD, in which the first
condition in Eq.\ref{FD2} is irrelevant. As in CSD, we shall continue to assume the
second condition in Eq.\ref{FD2} is accurately maintained, while we shall allow
a small violation of the third condition in Eq.\ref{FD2} parametrized by a small
parameter $\varepsilon$,
\begin{equation}
\label{PCSD}
B = \frac{b}{\sqrt{3}}
\left(
\begin{array}{r}
1 \\
1 \\
-1
\end{array}
\right),\ \
C  = \frac{c}{\sqrt{2}}
\left(
\begin{array}{r}
\varepsilon \\
1 \\
1
\end{array}
\right).
\end{equation}
We refer to this as Partially Constrained Sequential Dominance (PCSD), since
one of the conditions of CSD is maintained, while the other one is violated by the
parameter $\varepsilon$. Note that the introduction of the parameter $\varepsilon$
also implies a violation of FD since the columns of the Dirac mass matrix
$B,C$ can no longer be identified with the columns of the MNS matrix, due to the
non-orthogonality of $B$ and $C$. In other words the
$R$ matrix, the general orthogonal matrix introduced by Casas and
Ibarra \cite{Casas:2001sr}, will differ from the unit matrix by an amount of order
$\varepsilon$, allowing non-zero leptogenesis as discussed in
\cite{King:2006hn,Antusch:2006cw}.

Assuming PCSD as in Eq.\ref{PCSD}, with a strong neutrino mass hierarchy
$|m_1|\ll |m_2| < |m_3|$, leads to a neutrino mass matrix,
\begin{equation}
\label{PCSD2}
({M^{\nu}_{eff}})^{PCSD} = \frac{BB^{T}}{M_{B}} +\frac{CC^{T}}{M_{C}},
\end{equation}
where
\bea\label{PCSD3}
B B^{T} & = &  \frac{b^2}{3}
\left(\begin{array}{rrr}
1 & 1 & -1 \\
1 & 1 & -1 \\
-1 & -1 & 1
\end{array}\right), \nonumber \\
C C^T & = &  \frac{c^2}{2}\left(\begin{array}{rrr}
0 & 0 & 0 \\
0 & 1 & 1 \\
0 & 1 & 1
\end{array}\right)
+ \varepsilon\frac{c^2}{2}
\left(\begin{array}{rrr}
0 & 1 & 1 \\
1 & 0 & 0 \\
1 & 0 & 0
\end{array}\right).
\eea
To leading order in $|m_2|/|m_3|$ the mass matrix $({M^{\nu}_{eff}})^{PCSD}$
in Eqs.\ref{PCSD2},\ref{PCSD3} corresponds to $({M^{\nu}_{eff}})^{TBR}$
in Eqs.\ref{MTBR},\ref{MTBR2}
where we identify,
\beq
m_1=0, \ \ m_2=b^2/M_B, \ \ m_3=c^2/M_C, \ \ \varepsilon = re^{-i\delta }.
\eeq
Thus, the TBR form of mixing matrix in Eq.\ref{MNS3} will result, to leading order
in $|m_2|/|m_3|$.

It is straightforward to implement the above example of PCSD into realistic GUT models
with non-Abelian
family symmetry spontaneously broken by flavons which are based on the CSD
mechanism \cite{King:2005bj,deMedeirosVarzielas:2005ax,King:2006me}.
In such models the columns of the Dirac mass matrix in the diagonal
right-handed neutrino mass basis are determined by flavon vacuum alignment,
with the column $B$ identified with a triplet flavon $\phi_{123}$ and
the column $C$ identified with a triplet flavon $\phi_{23}$
and it is quite easy to obtain a correction to the vacuum aligmment
such that
\begin{equation}
\label{PCSD4}
\langle \phi_{123} \rangle  = \frac{b}{\sqrt{3}}
\left(
\begin{array}{r}
1 \\
1 \\
-1
\end{array}
\right),\ \
\langle \phi_{23} \rangle  = \frac{c}{\sqrt{2}}
\left(
\begin{array}{r}
\varepsilon \\
1 \\
1
\end{array}
\right),
\end{equation}
in direct correspondence with Eq.\ref{PCSD}.
For example, in such models based on the discrete family symmetry $A_4$
\cite{King:2006me},
the flavon vacuum expectation
value (VEV) $\langle \phi_{123} \rangle $ will preserve a $Z_2$
subgroup of the original discrete family symmetry corresponding to an $A_4$
generator $S$ \cite{Altarelli:2006kg}, while the flavon VEV $\langle \phi_{23} \rangle $ will
violate this subgroup even in the limit that $\varepsilon=0$. It is therefore
natural to assume some misalignment of $\langle \phi_{23} \rangle $ since,
unlike $\langle \phi_{123} \rangle $, it is not protected by any symmetry.

To be concrete, in the $A_4$ Pati-Salam model in \cite{King:2006me}
a radiative symmetry breaking mechanism is used, and in the first stage of symmetry
breaking the alignment is achieved via an $A_4$
invariant term but $SO(3)$-breaking term $\lambda \phi^{\dagger i}\phi^{i}
\phi^{\dagger i}\phi^{i}$ (summed over $i=1,2,3$) which aligns the VEV
of the field $\phi$ in the following possible directions,
\be
\label{available-vacua} \langle|\phi|\rangle\propto (1,1,1)\quad
{\rm and/or}\quad \langle|\phi|\rangle\propto
(1,0,0),\,(0,1,0),\,(0,0,1)
\ee
where only the magnitudes of the components are so far specified.
What matters is the sign of the
$SO(3)$-breaking term: if $\lambda > 0$ the ``isotropic'' option
$\langle | \phi |\rangle\propto (1,1,1)$ is picked up while the VEV is
maximally ``anisotropic'' (i.e. with just one nonzero entry in
$\langle\phi\rangle$) if $\lambda <0 $. Following this approach,
in the first stage, triplet flavons $\phi_{123}$, $\phi_{1}$ and $\phi_{3}$ are introduced
and the above mechanism is used to align the magnitudes of the VEVs as
\be\label{first-stage}
\langle|\phi_{123}|\rangle\propto (1,1,1)\quad {\rm and}
\quad \langle|\phi_{1}|\rangle\propto (1,0,0),\quad \langle|\phi_{3}|\rangle\propto (0,0,1).
\ee
In the second stage, two further triplet flavons $\phi_{23}$, $\tilde{\phi}_{23}$ were introduced \cite{King:2006me}
and a potential was assumed to lead to the CSD alignment. Here we show how a small modification
of this potential can lead to alignment along the direction in Eq.\ref{PCSD4}.
We continue to assume that their VEVs are radiatively driven, but assume the slightly different alignment terms
\be\label{SO3potential}
\tilde{\lambda}_{123}|\phi_{123}^{\dagger}.\tilde{\phi}_{23}|^{2}+\tilde{\lambda}_{1}|\phi_{1}^{\dagger}.\tilde{\phi}_{23}|^{2}
+ \lambda_{1}|\phi_{1}^{\dagger}.{\phi}_{23}|^{2} + \lambda_{23}|\tilde{\phi}_{23}^{\dagger}.{\phi}_{23}|^{2}.
\ee
If $\tilde{\lambda}_{123}$ is positive, the VEV of
$\tilde{\phi}_{23}$ is driven to be orthogonal to
$\langle{\phi}_{123}\rangle$ while $\tilde\lambda_{1}>0$
makes its first component vanish and thus
$\langle|{\tilde\phi}_{23}|\rangle\propto
(0,1,1)$. In order to obtain the alignment of $\langle{\phi}_{23}\rangle$ in Eq.\ref{PCSD4}
we now assume negative $\lambda_{23}$, which tends to align $\langle{\phi}_{23}\rangle$ along the direction
of $\langle\tilde{\phi}_{23}\rangle$, and also assume that $\lambda_1$ is negative, which will
tend to align $\langle{\phi}_{23}\rangle$ along the direction
of $\langle {\phi}_{1}\rangle$. The combined effect of these two terms will be to lead to
an alignment of $\langle{\phi}_{23}\rangle$ of the form assumed in Eq.\ref{PCSD4}.
The message is that, since no symmetry is determining the alignment of
$\langle{\phi}_{23}\rangle$, the choice of vacuum alignment for this flavon is
not protected, and slightly different dynamical assumptions will lead to slightly different
alignments. By contrast the alignment of  $\langle \phi_{123} \rangle $ is determined
by simpler dynamics driven by the $A_4$ symmetry.

In such realistic GUT models there will also be additional corrections from
TBR mixing due to charged lepton corrections, renormalization group (RG) running,
and canonical normalization (CN) effects, as fully studied in \cite{Antusch:2008yc}.
In such a realistic framework, the TBR mixing matrix described here would correspond
to the leading order neutrino mixing matrix at the GUT scale, with modified
neutrino mixing sum rules as discussed in \cite{Antusch:2008yc}.
In such a GUT-flavour framework, one expects the
charged lepton corrections to the neutrino mixing angles to be less than of order
$\theta_{12}^e/\sqrt{2}$
(where typically $\theta_{12}^e$ is a third of the Cabibbo angle)
plus perhaps a further $1^o$ from RG corrections.
Thus such theoretical corrections cannot account for an observed reactor angle
as large as $8^o$, starting from the hypothesis of exact TB neutrino mixing.

In conclusion, we have proposed an extension of tri-bimaximal mixing to include a
non-zero reactor angle $\theta_{13}$ while maintaining the tri-bimaximal predictions for the
atmospheric angle $\theta_{23}=45^o$ and solar angle $\theta_{12}=35^o$.
The TBR mixing proposal, which, to leading order in $r$, predicts the deviation parameters $s=a=0$ for
all $r$, is distinct from the tri-maximal mixing proposal which, to the same approximation, predicts
$s=0$ but $a=-(r/2)\cos \delta$.
We have shown how such tri-bimaximal-reactor mixing can arise at leading order
from the (type I) see-saw mechanism with partially constrained sequential dominance.
We have in turn shown how partially constrained sequential dominance
can be realized in GUT models with a non-Abelian discrete
family symmetry, such as $A_4$, spontaneously broken by flavons with a
particular vacuum alignment.
In such models there will be theoretical corrections to the TBR mixing matrix
due to charged lepton corrections, RG running and CN effects, but these are
all expected to be subdominant compared to a large reactor angle $\theta_{13}$
in the region of $8^o$. Indeed, if RG/CN effects are very small
and $\theta_{12}^e$ is much smaller than the Cabibbo angle,
then the tri-bimaximal-reactor predictions predictions for the
atmospheric angle $\theta_{23}=45^o$ and solar angle $\theta_{12}=35^o$ would be
realized very precisely. If the present indication for $\theta_{13}$ holds up,
it will be very interesting to see how close the solar and atmospheric angles
are to their tri-bimaximal values.

\section*{Acknowledgements}
We would like to thank the organizers of the
XIII International Workshop on
"Neutrino Telescopes", Venice, March 10-13, 2009, Venice
for their kind hospitality, and for partial
support from the following grants: STFC Rolling Grant
ST/G000557/1; EU Network MRTN-CT-2004-503369; EU ILIAS RII3-CT-2004-506222.

\end{document}